\documentclass[12pt]{iopart}

\usepackage{graphicx}
\usepackage{url}
\usepackage{hyperref}
\usepackage{siunitx}
\usepackage{color}

\begin{document}

\title{Cold neutral atoms via charge exchange from excited state positronium: a proposal}

\author{W. A. Bertsche}

\address{School of Physics and Astronomy, University of Manchester, Manchester M13 9PL, UK \\
The Cockcroft Institute, Daresbury Laboratory, Warrington WA4 4AD, UK}


\author{M. Charlton and S. Eriksson}
\ead{s.j.eriksson@swansea.ac.uk}

\address{ Department of Physics, College of Science, Swansea University, Singleton Park, Swansea SA2 8PP, UK}


\date{\today}

\begin{abstract}

We present a method for generating cold neutral atoms  via charge exchange reactions between trapped ions and Rydberg positronium. The high charge exchange reaction cross section leads to efficient neutralisation of the ions and since the positronium-ion mass ratio is small, the neutrals do not gain appreciable kinetic energy in the process. When the original ions are cold the reaction produces  neutrals that can be trapped or further manipulated with electromagnetic fields. Because a wide range of species can be targeted we envisage that our scheme may enable experiments at low temperature that have been hitherto intractable due to a lack of cooling methods. We present an estimate for achievable temperatures, neutral number and density in an experiment where the neutrals are formed at a milli-Kelvin temperature from either directly or sympathetically cooled ions confined on an ion chip. The neutrals may then be confined by their magnetic moment in a co-located magnetic minimum well also formed on the chip. We discuss general experimental requirements. 
\end{abstract}


\maketitle

\section{Introduction}

Low temperature and charge neutrality are cornerstones in atomic physics, allowing investigations of phenomena of both fundamental and applied importance. Historically, a growing number of experimentally available cold neutral species has been inextricably linked with new research. Examples include using ultracold alkaline earth-like elements in optical frequency standards~\cite{Ludlow2015}, and cooling chromium~\cite{Griesmaier2005} and rare earth elements~\cite{Mcclelland2006} with a large magnetic moment to access novel interactions. Some halogens can be cooled~\cite{Rennick2014} and with a growing number of cold species including molecules~\cite{Carr2009}, chemistry at low energy is an emerging field. Laser cooling is the paradigm for achieving low temperatures, but only in select atomic species and recently in certain classes of molecules~\cite{Norrgard2016} with energy transitions that match available laser sources. Despite technological advances, direct laser cooling of universally abundant species such as hydrogen, carbon and oxygen~\cite{Tielens2013}, as well as molecules in general, remains either intractable or extremely challenging at best. Producing cold molecules via photo-association or Feshbach resonances in laser cooled species faces similar limitations. Other generally applicable methods like buffer gas cooling and deceleration in switched electromagnetic fields produce either high density or low temperatures but often not both. Very recently, ultracold polar molecules have been achieved by optoelectrical cooling~\cite{Prehn2016}.  A general method for producing arbitrary cold neutral species at or below milli-Kelvin temperatures is, however, still outstanding.

Here we describe how excited state positronium (Ps) can be used to produce cold neutral atoms or molecules (X) from target ions ($\rm{X^+}$) by the reaction
\begin{equation}
  \rm{X^+ + Ps \rightarrow X + e^+}.
\label{eq:exchange}
\end{equation}

The reaction is initiated by exciting the Ps to a high-lying Rydberg state after the target ion has been confined in a Paul trap and cooled either by direct laser cooling or coupling to a thermal bath. The latter can be achieved by means such as sympathetic cooling with another laser-coolable ion, or by coupling the ion motion to a cryogenically cooled resistor~\cite{Itano95}. The latter technique becomes relevant for (anti)protons. The advantage of Ps is that, as detailed below, the recoil energy imparted in the collision is small enough that the increase in the kinetic energy of the trapped neutral is negligible, such that it will be formed at or near the temperature of the original ion. After the neutral atom or molecule forms it may be confined by interaction with its magnetic moment in a co-located magnetic minimum well where its electronic excitation decays, or is controllably removed by a stimulated Raman process. Once the atom is in a known state it can be detected by interaction with radiation or further manipulated in shallow potentials made by electromagnetic fields. We introduce a hybrid atom/ion chip~\cite{Stick2005, Fortagh2007} which allows convenient integration of the central elements. In principle the concept applies to any atomic or molecular species that can be cooled in an ion trap, as the strength of the Coulomb interaction eases the initial trapping of the target ion.

\section{Charge exchange reactions with excited state positronium}

Recent years have seen major advances in the production and manipulation of positronium in vacuum. 
Ps can be produced efficiently at low kinetic energies using a porous silica sample into which a \SI{}{\kilo\electronvolt}-energy positron beam is implanted. The Ps atoms are emitted from the surface of the pores with around \SI{1}{\electronvolt} of kinetic energy and are then cooled by collision as they pass through the pore volume, before being emitted into vacuum. Kinetic energies in the \numrange[range-phrase= --]{25}{70} \SI{}{\milli\electronvolt} range have been deduced \cite{Crivelli10,Cassidy10,Deller15}, dependent upon details of the sample structure.
Importantly, a  number of groups currently have the capability to produce excited Ps atoms across a wide range of principal quantum numbers $n_{\rm{Ps}}$ \cite{Deller15,Jones14,Aghion16,Wall15,Jones2016,Baker16}, with efficiencies of around 20\% of the incident positron flux, using pulsed laser systems timed to coincide with bursts of positrons ejected from buffer gas-type accumulation devices (e.g., \cite{Clarke06,Cassidy06}). 

Our study of the Ps-X$^+$ reaction has been motivated in part by a recent resurgence of interest in the Ps-p system, and in particular the charge exchange reaction Ps + p $\rightarrow$ H + e$^+$. The charge conjugate of this process has long been postulated as a candidate for direct antihydrogen formation (\cite{Humb87,Charlton90}), and two groups are close to implementation (see e.g., \cite{vdWerf14,Krasnicky14}). In addition to production by directly mixing positrons and antiprotons ~\cite{ALPHATrap1,Enomoto2010,Gabrielse2012}, antihydrogen has also been produced in a double Rydberg charge exchange scheme that involved the creation and interaction of Ps~\cite{Storry2004} with recent improvements in Ps yield~\cite{McConnell2016}. Recent, high quality, calculations for Ps-p charge exchange \cite{Kadyrov15,Rawlins16} for initial Ps states from $n_{\rm{Ps}}$= 1-3 have revealed unexpectedly large cross sections that increase dramatically with $n_{\rm{Ps}}$ and (other than for the ground state) exhibit a $1/K_{\rm{Ps}}$ behaviour at low Ps kinetic energies, $K_{\rm{Ps}}$. This work has been supplemented by a Classical Trajectory Monte Carlo (CTMC) investigation explicitly for Rydbergs \cite{Krasnicky2016}, and these studies are used to provide estimates of required Ps fluxes in the next section.

In the present work it is assumed that the Ps-p system is typical of the Ps-ion class which is characterised by a long-range charge-dipole interaction potential which ensures that the ion interacts with the Rydberg Ps as a whole \cite{Gallagher,Gallagher1988}. As such, the collision can effectively be broken down into separate $\rm{e}^-$-$\rm{X}^+$ and $\rm{e}^+$-$\rm{X}^+$ systems which will be dominated by the former, leading to charge transfer, due to the repulsive nature of the positron-ion interaction. In effect, at the collision energies we consider, the orientation of the $\rm{e}^-$-$\rm{e}^+$ dipole is frozen relative to the ion, irrespective of species, resulting in broadly similar scattering cross sections for all singly charged ions. We also assume that the cross sections for reaction (\ref{eq:exchange}) ensure that the charge transfer process remains competitive with other collisional outcomes, such as break-up ($\rm{X^+} + \rm{Ps} \rightarrow \rm{X + e^+ + e^-}$) and radiative attachment (${\rm{X^+ + Ps \rightarrow X^+ Ps }}$ $ + $ $  h\nu$), for a wide variety of species. We note, though, that little is known concerning very low energy Ps-X$^+$ interactions involving excited states of Ps, other than the aforementioned work on the Ps-p system.

One outcome of the Ps-ion interaction is a Rydberg neutral in a state in which the binding energy difference, $Q$, between it and the starting Ps state is close to zero, such that the collision is quasi-resonant (see \cite{Massey74} for a discussion of this phenomenon in charge transfer reactions). It is easy to show that the kinetic energy imparted to the atomic species (of mass $m_{\rm{X}}$) as a result of $Q$ is of the order of $Q m_e / m_{\rm{X}}$, with $m_e$ the electron mass. Thus, this recoil can be neglected for Rydberg Ps atoms, which can have binding energies in the region of \SI{10}{\milli\electronvolt}. Reaction (\ref{eq:exchange}) is likely to be very kinematically forward-peaked, thus the kinetic energy gain will be around the Ps kinetic energy, $K_{\rm{Ps}}$, suppressed by the factor $m_e/m_{\rm{X}}$. Using this, and requiring that the resulting neutral kinetic energy must be less than the depth of the atom trap (given by $3k_BT_W/2$, with $T_W$ the effective trap wall temperature), one can show that $K_{\rm{Ps}}$ must be less than approximately $0.24T_WM_{\rm{X}}$ (\SI{}{\electronvolt}, with $T_W$ in kelvin), where $M_{\rm{X}}$ is the mass number of the trapped atom. The low reaction recoil, together with the low momentum brought into the collision by the Ps, results in the total recoil energy of the neutrals being small enough for them to be held in shallow magnetic minimum traps. With $T_W$ around \SI{10}{\milli\kelvin} for a magnetic trap for the atom (see section 4), the required maximum $K_{\rm{Ps}}$ for  $M_{\rm{X}} \ge 10$ is already near the capabilities of demonstrated porous silica Ps targets that produce a large fraction of Ps with kinetic energies close to room temperature (see section 3). Thus, a wide variety of species created via Ps charge exchange can be held, provided cold enough target ions are available.

A further feature of reaction (\ref{eq:exchange}) is that the final Rydberg state of the target atom is controlled entirely by that of the Ps, since the ejected positron is essentially a spectator. This is in contrast to the case of a charge exchange reaction between a target ion and a heavier neutral atom (which itself would have to be cold beforehand to prevent excessive heating on collision) for which a slew of final states of both the newly formed neutral and the resulting ionised atom would be possible.

\section{Estimate of required positronium flux}

Whilst reaction (\ref{eq:exchange}) has advantages over charge exchange involving massive species, a potential disadvantage is that Ps is relatively scarce and must be produced in-situ using a low energy positron beam. Thus, we provide an estimate of the rate of cold neutral production via Rydberg Ps-ion interactions to further assess the viability of the method and estimate the positron beam requirements for this technique. It can be shown that the number of Ps emitted isotropically from a point source required to neutralise a fraction $f$ of ions located in a trap at a distance $R$ away is approximately $ \frac{2 \pi R^2}{\sigma} f$, where $\sigma$ is the scattering cross section. We estimate $\sigma$ using guidance from the p-Ps results of Bray and Kadyrov and co-workers \cite{Kadyrov15,Rawlins16}, whose (anti)hydrogen production data show, for example, an order of magnitude increase in the cross section between Ps(2s) and Ps(3s), with the latter close to \SI{1.5e-16}{\square\meter} for $K_{\rm{Ps}}$ of \SI{10}{\milli\electronvolt}. Recent CTMC results \cite{Krasnicky2016} are in remarkable accord with the Bray/Kadyrov data where they overlap for $n_{\rm Ps} = 3$, giving confidence  that the former can provide useful estimates of the cross sections for Rydberg Ps. The CTMC study finds that the charge transfer cross sections scale as $n_{\rm Ps}^4$, as expected from a Bohr analysis, such that $\sigma$ reaches around \SI{e-12}{\square\meter} with $n_{\rm{Ps}} \approx 50$ at a Ps kinetic energy of about 10 \SI{}{\milli\electronvolt}. With Ps emanating from a silica target located \SI{e-2}{\meter} away (as illustrated in Fig. \ref{fig:concept}, see section 4 for details), around \num{6e8}  Rydberg Ps will be needed to  neutralise all ions.

Porous silica targets offer excellent conversion efficiency from positrons impacting on the surface to reflected positronium that cools towards its target temperature \cite{Ferragut2013,Andersen2015}. In our example, taking the aforementioned positron-Rydberg conversion efficiency of 20\%, approximately \num{3e9} positrons would be needed to fully neutralise the ion sample, and this scales with incoming positron flux for applications where complete conversion is not required. It has recently been demonstrated that \num{4e9} positrons can be stored in an accumulator~\cite{Fitzakerley2016}. Whilst this is, in principle, sufficient to ensure that $f =1$ using our simple approximation, we note that: (i) further gains in positron number are feasible using laboratory-based positron systems with extra storage capabilities~\cite{Surko03}, (ii) sources with significantly higher positron fluxes are under development~\cite{Perez2015} and (iii) higher fluxes are available at existing facility based sources~\cite{Hugenschmidt2014}. 

The arrival of such a large pulse of charge may lead to ion heating in the \SI{}{\micro\electronvolt} regime which can be mitigated by shielding the positron beam or splitting up the pulse into shorter bunches. Alternatively, the Ps converter can be back illuminated with some loss of positrons: e.g., work in \cite{Andersen2015} shows conversion efficiencies on transmission through a carbon-backed porous silica membrane of $\sim10\%$. Though the efficiency of conversion is lower, integrating such a target directly into the chip ion trap would decrease the distance $R$ from converter to ions and hence lower the positron requirements significantly (by as much as a factor of 20), making the technique immediately viable for positron systems that have existed for well over a decade~\cite{Jorgensen05}. We also envisage that several shots of neutralised atoms could be collected in the magnetic trap over time by simultaneously operating the two traps and re-loading ions as they are depleted. 

\section{A concept for neutral atom production and trapping}

The formed neutral can be trapped in a magnetic minimum potential well provided that the depth is commensurate with the original ion energy and that it has a long-lived state with a magnetic moment.  Magnetic trap field gradients of around $b'=\SI{5}{\tesla\per\meter}$ are technically feasible in conventional magnetic traps even without superconductors and of order \SI{10}{\milli\kelvin} depth is easily available for an atom with a magnetic moment $\mu=\mu_B$. The electronic excitation of the trapped target neutral decays in a time-scale much shorter than the typical magnetic trap lifetimes, which exceed \SI{10}{\second}. We note that the currently demonstrated route to trapped ground state antihydrogen relies on decay from the excited state in which the anti-atom is born \cite{ALPHATrap1}. Indeed, one can show that from a distribution of (anti)hydrogen atoms in random magnetic states up to $n_{\rm{H}}$=20, 50\% decay by a radiative cascade to the ground state in less than \SI{0.1}{\milli\second}~\cite{Topcu2006}, and at low magnetic fields on average one in four of these will be in a trapped state. Thus, even without any internal state control such as optical pumping, starting from  \num{e3} neutralised ions a few \num{e2} neutrals could be trapped per positron bunch. Assuming a uniform distribution of $\mathcal{N}$ atoms in a spherical volume, and that atoms will explore a radius up to $k_BT/\mu b'$  an estimate of the density can be given as $N=3(\mu b')^3 \mathcal{N}/4\pi (k_B T)^3$.  For example, with $b'=\SI{5}{\tesla\per\meter}$ and $T=\SI{5}{\milli\kelvin}$ one achieves $N=\SI{7e7}{\per\meter\cubed}$ per atom with $\mu=\mu_B$. Thus, assuming that 25\% of $\mathcal{N}=\num{1.0e3}$ are trapped in the ground state as in the hydrogenic case, we get a conservative estimate of $N=\SI{1.8 e10}{\per\cubic\meter}$.  We note that there is significant scope for improvement for atoms atom with higher magnetic moments. Alternatively, atom chip traps can provide somewhat higher field gradients and superconducting traps can provide significantly higher field gradients. For example, with $b'=\SI{220}{\tesla\per\meter}$ as in ref.~\cite{vanRoijen1988}, and with the other parameters as above, we would have $N=\SI{1.5 e15}{\per\cubic\meter}$, although in such an environment care must be taken to ensure that the Ps survives to interaction. 

\begin{figure}
\begin{center}
\includegraphics[width=0.7\linewidth]{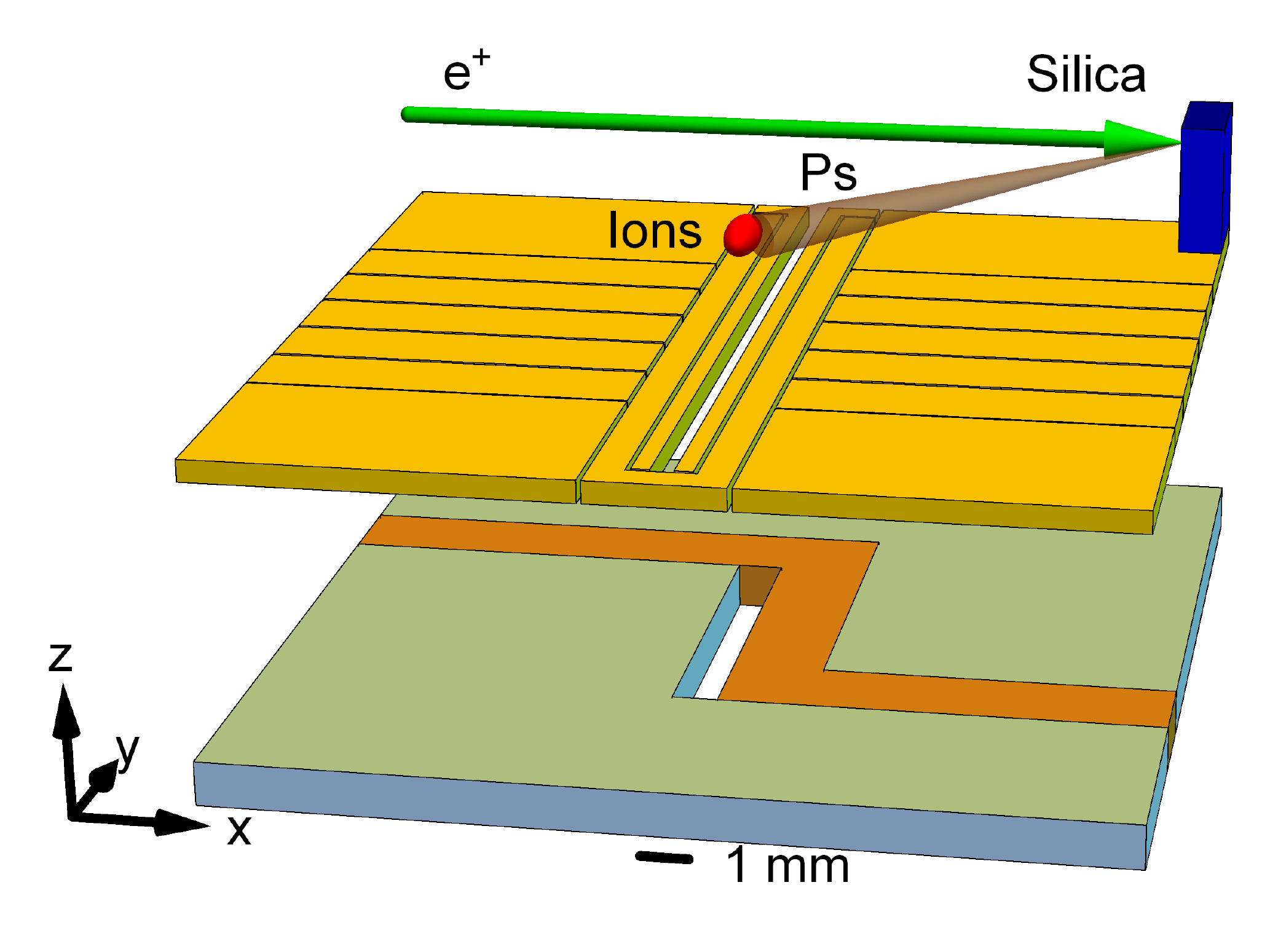}
\caption{\label{fig:concept} A hybrid ion-atom chip, including a Ps source. The layer that creates the magnetic trap for neutral atoms is offset along $z$ and appears below the ion trap chip for clarity; the latter is located directly above the magnetic trap layer. See text for details.}
\end{center}
\end{figure}

\begin{figure}
\begin{center}
\includegraphics[width=0.7\linewidth]{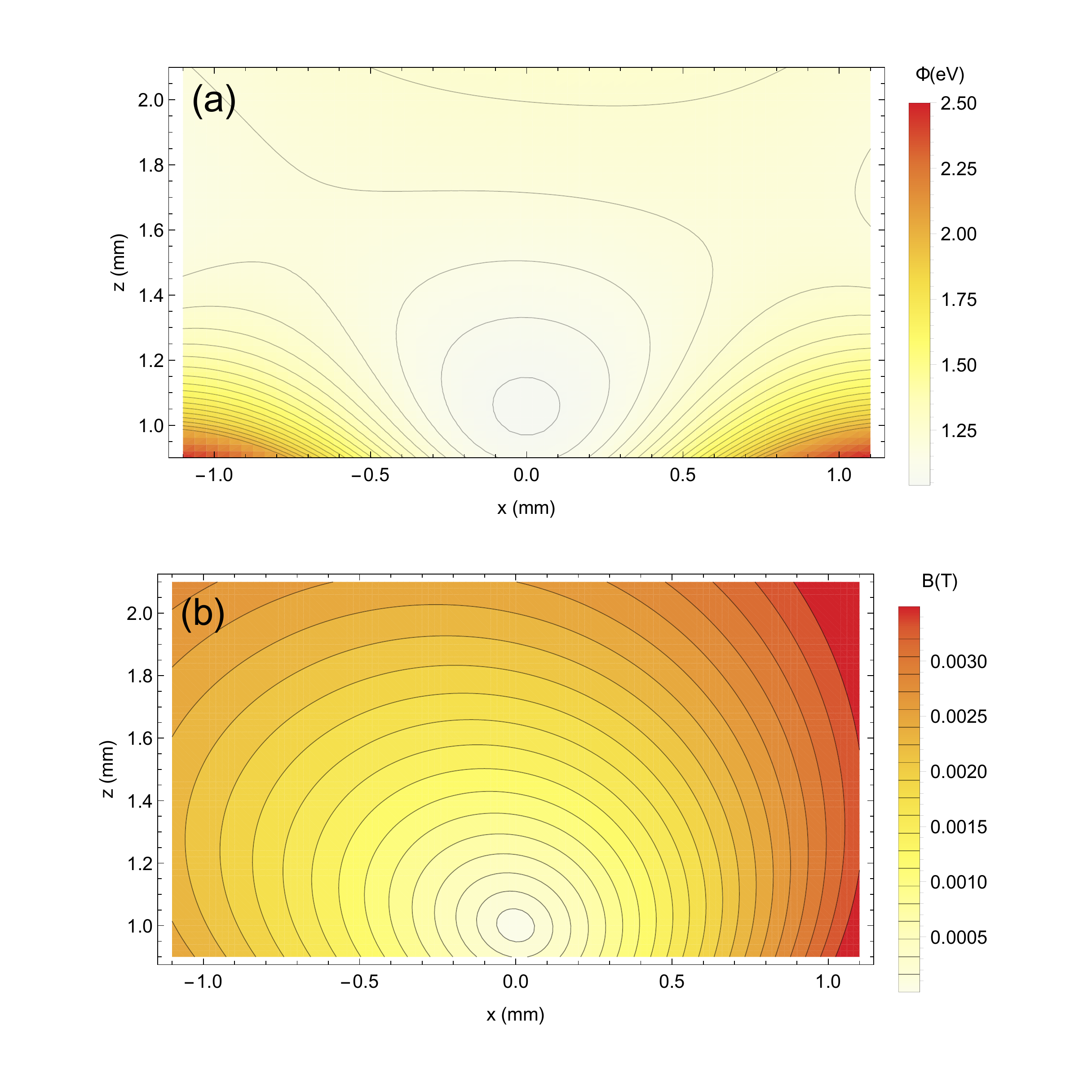}
\caption{\label{fig:bothpotentials} Cross sections in the $xz$-plane of (a) the combined RF and DC potential for Ca$^+$ ions for typical trapping parameters (see text) and (b) a neutral atom magnetic trap produced with 100 A in the wire and a 6.7 mT bias field creates a minimum which overlaps the ion trap minimum. The contour interval in (a) is \SI{50}{\milli\electronvolt} and in (b) there are 20 contours in the interval 0-3.5  \SI{}{\milli\tesla}}
\end{center}
\end{figure}

Our concept can be realised by superimposing a conventional Paul trap for ions and a conventional trap for neutral atoms for which typical trapping parameters were given above. We envisage that a hybrid atom/ion chip geometry as shown in Fig.~\ref{fig:concept} 
offers simplification and further advantages (see below) and thus we now turn to discuss this specific scenario in more detail. In Fig.~\ref{fig:concept}, trapped ions (red)  are neutralised by Ps (brown) originating from the silica source (dark blue). Both the incident positron beam (green) used for Ps production and the silica source are separated by $\sim$ \SI{e-2}{\meter} from the ions to reduce electrostatic interference and to allow access for laser cooling and detection by standard ion trapping methods together with the Ps excitation laser beam (not shown). The ion-chip electrodes are connected as described in~\cite{Allcock2010}, but the physical dimensions have been chosen to create a trap about \SI{1}{\milli\meter} from the surface and a chip layer thickness of \SI{0.5}{\milli\meter} allows fabrication from foil~\cite{Mokhberi} for convenience. The central DC-electrodes are separated by a \SI{0.5}{\milli\meter} gap to allow loading the ions from below the surface as described in~\cite{Mokhberi}. The internal separation and width of the RF electrodes are \SI{1.5}{\milli\meter} and \SI{3.0}{\milli\meter}  respectively.  With an RF excitation voltage of \SI{400}{\volt} at \SI{8}{\mega\hertz}, voltages ranging from \SIrange{-2}{+3}{\volt} on the central DC electrodes, and around \SI{17}{\volt} on the end-caps, we calculate using~\cite{Schmied2010,Schmied:package} that the ion trap depth becomes \SI{160}{\milli\electronvolt} for Ca$^+$ ions. These parameters yield trapping frequencies of a few \SI{100}{\kilo\hertz}. The combined ion potential in the centre of the chip is shown in Fig.~\ref{fig:bothpotentials}~(a).
These parameters, while somewhat arbitrary, illustrate typical properties of a trapping potential with minimum energy far enough away from the surface that heating due to the influence of patch-effects is reduced. Further, the DC-control voltages are chosen to bring the combined potential close to the RF-null line in order to minimise excess micromotion and to allow a slight tilt of the principal axes to facilitate laser cooling of all degrees of freedom with a laser beam parallel to the $xy$-plane~\cite{Allcock2010}. While constituting a realistic set of trapping parameters which enables trapping of order \num{e3} ions and forming Coulomb crystals at temperatures near \SI{5}{\milli\kelvin}, we note that further optimisation may be required in an experiment. We also note that very large numbers of ions are in principle available, e.g., $9\times\num{e4}$ ions have been reported~\cite{Removille2009}.

The magnetic potential well (Fig.~\ref{fig:bothpotentials}~(b)) with a minimum coinciding with the ion trap is generated by a single Z-shaped wire~\cite{Fortagh2007} slightly offset from the centre (to allow placing the ion source beneath the chip) together with a uniform bias field at an angle from the $xy$-plane to compensate for the offset. The centre wire segment creates the field gradient for atom confinement above the chip following a line along $y$ while the ends of the Z-wire pinch off the trap. The length of the trap along $y$ is determined by the length of the central segment. The position of the trap above the surface can be moved by changing the bias field or the wire current. This geometry allows an overall trap depth of about \SI{0.05}{\milli\kelvin\per\ampere} for an atom with a magnetic moment $\mu_B$. In order to trap most of the neutral atoms that form by charge exchange described here, of order \SI{100}{\ampere} needs to run in the copper Z-wire, achievable in a room temperature trap~\cite{Wildermuth2004}. Since field gradient in the chip trap is not isotropic the expression for density is no longer valid but nevertheless provides a guide to achievable neutral atom densities. However, since the field gradient in the chip trap  scales with the inverse distance squared~\cite{Fortagh2007}, it is not unreasonable to anticipate densities up to \SI{e12}{\per\cubic\meter} with small changes to the geometry. Further gain in trap depth and atom density could be achieved by making the ion-chip layer thinner, thus bringing the traps closer to the wire, and increasing the current density in the wire (e.g., by microfabrication). Alternatively, higher current densities could be achieved with a superconducting wire in a cryogenic system. The presence of the chip surface limits the achievable magnetic trap depth and may also the affect the final temperature of the ions due to field inhomogeneities which become more significant at close proximity. Deeper and cleaner potentials are available if the concept is realised  by conventional macroscopic (non-chip) field sources, or alternatively one may choose a combination of ion or atom chip technology. However, the proposed geometry offers increased freedom for optical access. 

\section{Outlook}
\label{section:outlook}

Without modification or extension this technique applies directly to the creation of cold (even if not trapped) neutrals in any atomic species that has singly charged bound states which can be used for cooling in a trap either through direct laser cooling or sympathetically through the use of a secondary refrigerant ion with well known laser cooling characteristics such as Ca$^+$ or Be$^+$. A particularly notable example is neutral carbon which cannot be laser cooled in practice due to the lack of laser sources at the wavelength of the transition to the first excited state. Following our proposed method, C$^+$ would be sympathetically cooled in the ion trap, then, after neutralisation, become magnetically trapped in one of its long lived $^3P_1$ or $^3P_2$ states. In the dual ion configuration it may become necessary to use the spatial separation that occurs in ion trapping in order to avoid both species interacting with the Ps atoms unless cold reactions between the participants are explicitly desired in the experiment. A massive atom gains little kinetic energy from neutralisation and therefore the initial density of the neutrals will be more like the density of the trapped ions. To avoid multiple species interactions, it is straightforward to focus the Rydberg excitation laser to a spot size smaller than commonly observed ion spatial separation distances of \SIrange{10}{100}{\micro\meter} (see e.g. \cite{Mokhberi}), and thereby limit the region over which the reaction has appreciable cross section to only cover the target species.

In principle cold neutral molecules can be produced with this method equally well, however, here the internal molecular structure is likely to complicate trapping the molecule since decay channels from the excited state are not as straightforward. To mitigate this problem the excited state population may be transferred to a trapped state, e.g. with a stimulated Raman process. Furthermore, reactions involving X and X$^+$, which (with cooling present) are mostly benign for atoms,  may pose limitations for molecules. We note that a variant of reaction (\ref{eq:exchange}) involving Ps collisions with negatively charged ions may be a novel source of exotic bound states~ (see e.g., \cite{Drachman79}) akin to Ps-atom systems. Finally, we predict that with a colder Ps source or deeper magnetic trap, hydrogen or antihydrogen can be created by reaction (\ref{eq:exchange}) from cold protons or antiprotons on a cryogenic variant of our hybrid chip trap. This method could potentially lead to higher antihydrogen experiment rates as the miniaturisation allows much improved optical access compared to current technology.

In summary, we have presented a method to produce and trap cold neutral species via charge exchange with excited state positronium atoms on a hybrid ion/atom chip. Our predicted density of trapped neutrals is immediately competitive for experiments in novel collisions, fundamental tests and precision measurements~\cite{Carr2009} and has the advantage that at least in principle a wide range of species can be neutralised and trapped. Especially in the case of more massive species the untrapped neutrals inherit the trapped ion density with potential to address cold controlled chemistry. We have shown how an improved field gradient can also yield such high densities in the trap. Further optimisation for specific species may lead to a higher flux of cold neutrals than predicted here. Naturally, our method also applies to the regime where only a single neutral is produced. In the case of antihydrogen, our method would allow reducing the trap volume by several orders of magnitude which leads to increased interaction probability with laser beams and thereby increased statistical precision in measurements. The element of scalability and flexibility offered on atom chips  presents a possibility to create two (or more) traps for different species which could be merged in a controlled way. Since only a single refrigerant species and therefore only one laser cooling laser setup could be used for many target species, our method is a simplified route toward controlled inter-species interactions at the expense of requiring a Ps source.

\section{Acknowledgments}
The authors thank the UK EPSRC for supporting their antihydrogen research. SE gratefully acknowledges the Leverhulme Trust for his Research Fellowship.

\vspace{1cm}

\bibliography{references_resubmit_v5}

\providecommand{\newblock}{}
\begin{thebibliography}{10}
\expandafter\ifx\csname url\endcsname\relax
  \def\url#1{{\tt #1}}\fi
\expandafter\ifx\csname urlprefix\endcsname\relax\def\urlprefix{URL }\fi
\providecommand{\eprint}[2][]{\url{#2}}

\bibitem{Ludlow2015}
Ludlow A~D, Boyd M~M, Ye J, Peik E and Schmidt P~O 2015 {\em Rev. Mod. Phys.\/}
  {\bf 87} 637
  \urlprefix\url{http://link.aps.org/doi/10.1103/RevModPhys.87.637}

\bibitem{Griesmaier2005}
Griesmaier A, Werner J, Hensler S, Stuhler J and Pfau T 2005 {\em Phys. Rev.
  Lett.\/} {\bf 94} 160401
  \urlprefix\url{http://link.aps.org/doi/10.1103/PhysRevLett.94.160401}

\bibitem{Mcclelland2006}
McClelland J~J and Hanssen J~L 2006 {\em Phys Rev. Lett.\/} {\bf 96} 143005
  \urlprefix\url{http://link.aps.org/abstract/PRL/v96/e143005}

\bibitem{Rennick2014}
Rennick C~J, Lam J, Doherty W~G and Softley T~P 2014 {\em Phys. Rev. Lett.\/}
  {\bf 112}(2) 023002
  \urlprefix\url{http://link.aps.org/doi/10.1103/PhysRevLett.112.023002}

\bibitem{Carr2009}
Carr L~D, DeMille D, Krems R~V and Ye J 2009 {\em New J. Phys.\/} {\bf 11}
  055049 \urlprefix\url{http://stacks.iop.org/1367-2630/11/i=5/a=055049}

\bibitem{Norrgard2016}
Norrgard E~B, McCarron D~J, Steinecker M~H, Tarbutt M~R and DeMille D 2016 {\em
  Phys. Rev. Lett.\/} {\bf 116}(6) 063004
  \urlprefix\url{http://link.aps.org/doi/10.1103/PhysRevLett.116.063004}

\bibitem{Tielens2013}
Tielens A~G~G~M 2013 {\em Rev. Mod. Phys.\/} {\bf 85} 1021
  \urlprefix\url{http://link.aps.org/doi/10.1103/RevModPhys.85.1021}

\bibitem{Prehn2016}
Prehn A, Ibr\"ugger M, Gl\"ockner R, Rempe G and Zeppenfeld M 2016 {\em Phys.
  Rev. Lett.\/} {\bf 116} 063005
  \urlprefix\url{http://link.aps.org/doi/10.1103/PhysRevLett.116.063005}

\bibitem{Itano95}
Itano W~M, Bergquist J~C, Bollinger J~J and Wineland D~J 1995 {\em Phys. Scr.
  T\/} {\bf T59} 106
  \urlprefix\url{http://stacks.iop.org/1402-4896/1995/i=T59/a=013}

\bibitem{Stick2005}
Stick D, Hensinger W~K, Olmschenk S, Madsen M~J, Schwab K and Monroe C 2006
  {\em Nature Phys.\/} {\bf 2} 36
  \urlprefix\url{http://www.nature.com/nphys/journal/v2/n1/full/nphys171.html}

\bibitem{Fortagh2007}
Fortagh J and Zimmermann C 2007 {\em Rev. Mod. Phys.\/} {\bf 79} 235
  \urlprefix\url{http://link.aps.org/abstract/RMP/v79/p235}

\bibitem{Crivelli10}
Crivelli P, Gendotti U, Rubbia A, Liszkay L, P{\'e}rez P and Corbel C 2010 {\em
  Phys. Rev. A\/} {\bf 81} 052703
  \urlprefix\url{http://link.aps.org/doi/10.1103/PhysRevA.81.052703}

\bibitem{Cassidy10}
Cassidy D~B, Crivelli P, Hisakado T~H, Liszkay L, Meligne V~E, P\'{e}rez P, Tom
  H~W~K and {Mills Jr} A~P 2010 {\em Phys. Rev. A\/} {\bf 81} 012715
  \urlprefix\url{http://link.aps.org/doi/10.1103/PhysRevA.81.012715}

\bibitem{Deller15}
Deller A, Edwards D, Mortensen T, Isaac C~A, van~der Werf D~P, Telle H~H and
  Charlton M 2015 {\em J. Phys. B: At. Mol. Opt. Phys.\/} {\bf 48} 175001
  \urlprefix\url{http://stacks.iop.org/0953-4075/48/i=17/a=175001}

\bibitem{Jones14}
Jones A~C~L, Hisakado T~H, Goldman H~J, Tom H~W~K, {Mills Jr} A~P and Cassidy
  D~B 2014 {\em Phys. Rev. A\/} {\bf 90} 012503
  \urlprefix\url{http://link.aps.org/doi/10.1103/PhysRevA.90.012503}

\bibitem{Aghion16}
Aghion S {\em et~al.\/} (AEgIS Collaboration) 2016 {\em Phys. Rev. A\/} {\bf
  94} 012507 \urlprefix\url{http://link.aps.org/doi/10.1103/PhysRevA.94.012507}

\bibitem{Wall15}
Wall T~E, Alonso A~M, Cooper B~S, Deller A, Hogan S~D and Cassidy D~B 2015 {\em
  Phys. Rev. Lett.\/} {\bf 114} 173001
  \urlprefix\url{http://link.aps.org/doi/10.1103/PhysRevLett.114.173001}

\bibitem{Jones2016}
Jones A~C~L, Hisakado T~H, Goldman H~J, Tom H~W~K and {Mills Jr} A~P 2016 {\em
  Journal of Physics B: Atomic, Molecular and Optical Physics\/} {\bf 49}
  064006 \urlprefix\url{http://stacks.iop.org/0953-4075/49/i=6/a=064006}

\bibitem{Baker16}
Baker C~J, Edwards D, Isaac C~A, van~der Werf D~P and Charlton M 2017 {\em in
  preparation\/}

\bibitem{Clarke06}
Clarke J, van~der Werf D~P, Griffiths B, Beddows D~C~S, Charlton M, Telle H~H
  and Watkeys P~R 2006 {\em Rev. Sci. Instrum.\/} {\bf 77} 063302
  \urlprefix\url{http://scitation.aip.org/content/aip/journal/rsi/77/6/10.1063/1.2206561}

\bibitem{Cassidy06}
Cassidy D~B, Deng S~H~M, Greaves R~G and Mills~Jr A~P 2006 {\em Rev. Sci.
  Instrum.\/} {\bf 77} 073106
  \urlprefix\url{http://scitation.aip.org/content/aip/journal/rsi/77/7/10.1063/1.2221509}

\bibitem{Humb87}
Humberston J~W, Charlton M, Jacobsen F~M and Deutch B~I 1987 {\em J. Phys. B:
  At. Mol. Phys.\/} {\bf 20} L25
  \urlprefix\url{http://stacks.iop.org/0022-3700/20/i=1/a=005}

\bibitem{Charlton90}
Charlton M 1990 {\em Phys. Lett. A\/} {\bf 143} 143
  \urlprefix\url{http://www.sciencedirect.com/science/article/pii/037596019090665B}

\bibitem{vdWerf14}
van~der Werf D~P 2014 {\em Int. J. Mod. Phys.: Conference Series\/} {\bf 30}
  1460263
  \urlprefix\url{http://www.worldscientific.com/doi/abs/10.1142/S2010194514602634}

\bibitem{Krasnicky14}
Krasnick{\'y} D {\em et~al.\/} (AEgIS Collaboration) 2014 {\em Int. J. Mod.
  Phys.: Conference Series\/} {\bf 30} 1460262
  \urlprefix\url{http://www.worldscientific.com/doi/abs/10.1142/S2010194514602622}

\bibitem{ALPHATrap1}
Andresen G~B {\em et~al.\/} (ALPHA Collaboration) 2010 {\em Nature\/} {\bf 468}
  673
  \urlprefix\url{http://www.nature.com/nature/journal/v468/n7324/abs/nature09610.html}

\bibitem{Enomoto2010}
Enomoto Y, Kuroda N, Michishio K, Kim C~H, Higaki H, Nagata Y, Kanai Y, Torii
  H~A, Corradini M, Leali M, Lodi-Rizzini E, Mascagna V, Venturelli L, Zurlo N,
  Fujii K, Ohtsuka M, Tanaka K, Imao H, Nagashima Y, Matsuda Y, Juh\'asz B,
  Mohri A and Yamazaki Y 2010 {\em Phys. Rev. Lett.\/} {\bf 105} 243401
  \urlprefix\url{http://link.aps.org/doi/10.1103/PhysRevLett.105.243401}

\bibitem{Gabrielse2012}
Gabrielse G {\em et~al.\/} (ATRAP Collaboration) 2012 {\em Phys. Rev. Lett.\/}
  {\bf 108} 113002
  \urlprefix\url{http://link.aps.org/doi/10.1103/PhysRevLett.108.113002}

\bibitem{Storry2004}
Storry C~H {\em et~al.\/} (ATRAP Collaboration) 2004 {\em Phys. Rev. Lett.\/}
  {\bf 93} 263401
  \urlprefix\url{http://link.aps.org/doi/10.1103/PhysRevLett.93.263401}

\bibitem{McConnell2016}
McConnell R {\em et~al.\/} (ATRAP Collaboration) 2016 {\em J. Phys. B: At.,
  Mol. Opt. Phys.\/} {\bf 49} 064002
  \urlprefix\url{http://stacks.iop.org/0953-4075/49/i=6/a=064002}

\bibitem{Kadyrov15}
Kadyrov A~S, Rawlins C~M, Stelbovics A~T, Bray I and Charlton M 2015 {\em Phys.
  Rev. Lett.\/} {\bf 114} 183201
  \urlprefix\url{http://link.aps.org/doi/10.1103/PhysRevLett.114.183201}

\bibitem{Rawlins16}
Rawlins C~M, Kadyrov A~S, Stelbovics A~T, Bray I and Charlton M 2016 {\em Phys.
  Rev. A\/} {\bf 93} 012709
  \urlprefix\url{http://link.aps.org/doi/10.1103/PhysRevA.93.012709}

\bibitem{Krasnicky2016}
Krasnick\'y D, Caravita R, Canali C and Testera G 2016 {\em Phys. Rev. A\/}
  {\bf 94}(2) 022714
  \urlprefix\url{http://link.aps.org/doi/10.1103/PhysRevA.94.022714}

\bibitem{Gallagher}
Gallagher T~F 1994 {\em Rydberg Atoms\/} (Cambridge University Press,
  Cambridge)

\bibitem{Gallagher1988}
Gallagher T~F 1988 {\em Rep. Prog. Phys.\/} {\bf 51} 143
  \urlprefix\url{http://stacks.iop.org/0034-4885/51/i=2/a=001}

\bibitem{Massey74}
Massey H~S~W and Gilbody H~B 1974 {\em Electronic and Ionic Impact Phenomena
  Volume IV\/} (Clarendon, Oxford)

\bibitem{Ferragut2013}
Ferragut R, Aghion S, Tosi G, Consolati G, Quasso F, Longhi M, Galarneau A and
  Renzo F~D 2013 {\em J. Phys. Chem. C\/} {\bf 117} 26703
  \urlprefix\url{http://dx.doi.org/10.1021/jp410221m}

\bibitem{Andersen2015}
Andersen S~L, Cassidy D~B, Chevallier J, Cooper B~S, Deller A, Wall T~E and
  Uggerh{\o}j U~I 2015 {\em J. Phys. B: At., Mol. Opt. Phys.\/} {\bf 48} 204003
  \urlprefix\url{http://stacks.iop.org/0953-4075/48/i=20/a=204003}

\bibitem{Fitzakerley2016}
Fitzakerley D~W {\em et~al.\/} (ATRAP collaboration) 2016 {\em J. Phys. B: At.,
  Mol. Opt. Phys.\/} {\bf 49} 064001
  \urlprefix\url{http://stacks.iop.org/0953-4075/49/i=6/a=064001}

\bibitem{Surko03}
Surko C~M and Greaves R~G 2003 {\em Radiation Physics and Chemistry\/} {\bf 68}
  419--25 \urlprefix\url{http://dx.doi.org/10.1016/S0969-806X(03)00194-4}

\bibitem{Perez2015}
P{\'e}rez P {\em et~al.\/} (GBAR Collaboration) 2015 {\em Hyperfine
  Interactions\/} {\bf 233} 21--27 ISSN 1572-9540
  \urlprefix\url{http://dx.doi.org/10.1007/s10751-015-1154-8}

\bibitem{Hugenschmidt2014}
Hugenschmidt C, Ceeh H, Gigl T, Lippert F, Piochacz C, Reiner M, Schreckenbach
  K, Vohburger S, Weber J and Zimnik S 2014 {\em J. Phys.: Conference Series\/}
  {\bf 505} 012029
  \urlprefix\url{http://stacks.iop.org/1742-6596/505/i=1/a=012029}

\bibitem{Jorgensen05}
J\o{}rgensen L~V {\em et~al.\/} (ATHENA Collaboration) 2005 {\em Phys. Rev.
  Lett.\/} {\bf 95} 025002
  \urlprefix\url{http://link.aps.org/doi/10.1103/PhysRevLett.95.025002}

\bibitem{Topcu2006}
Top\ifmmode~\mbox{\c{c}}\else \c{c}\fi{}u T and Robicheaux F 2006 {\em Phys.
  Rev. A\/} {\bf 73} 043405
  \urlprefix\url{http://link.aps.org/doi/10.1103/PhysRevA.73.043405}

\bibitem{vanRoijen1988}
van Roijen R, Berkhout J~J, Jaakkola S and Walraven J~T~M 1988 {\em Phys. Rev.
  Lett.\/} {\bf 61}(8) 931--934
  \urlprefix\url{http://link.aps.org/doi/10.1103/PhysRevLett.61.931}

\bibitem{Allcock2010}
Allcock D~T~C, Sherman J~A, Stacey D~N, Burrell A~H, Curtis M~J, Imreh G, Linke
  N~M, Szwer D~J, Webster S~C, Steane A~M and Lucas D~M 2010 {\em New J.
  Phys.\/} {\bf 12} 053026
  \urlprefix\url{http://stacks.iop.org/1367-2630/12/i=5/a=053026}

\bibitem{Mokhberi}
Mokhberi A and Willitsch S 2014 {\em Phys. Rev. A\/} {\bf 90} 023402
  \urlprefix\url{http://link.aps.org/doi/10.1103/PhysRevA.90.023402}

\bibitem{Schmied2010}
Schmied R 2010 {\em New J. Phys.\/} {\bf 12} 023038
  \urlprefix\url{http://stacks.iop.org/1367-2630/12/i=2/a=023038}

\bibitem{Schmied:package}
Schmied R {\em SurfacePattern software package\/}
  \urlprefix\url{https://atom.physik.unibas.ch/people/romanschmied/code/SurfacePattern.php}

\bibitem{Removille2009}
Removille S, Dubessy R, Dubost B, Glorieux Q, Coudreau T, Guibal S, Likforman
  J~P and Guidoni L 2009 {\em Journal of Physics B: Atomic, Molecular and
  Optical Physics\/} {\bf 42} 154014
  \urlprefix\url{http://stacks.iop.org/0953-4075/42/i=15/a=154014}

\bibitem{Wildermuth2004}
Wildermuth S, Kr\"uger P, Becker C, Brajdic M, Haupt S, Kasper A, Folman R and
  Schmiedmayer J 2004 {\em Phys. Rev. A\/} {\bf 69} 030901
  \urlprefix\url{http://link.aps.org/doi/10.1103/PhysRevA.69.030901}

\bibitem{Drachman79}
Drachman R~J 1979 {\em Phys. Rev. A\/} {\bf 19} 1900
  \urlprefix\url{http://link.aps.org/doi/10.1103/PhysRevA.19.1900}

\end{thebibliography}

\end{document}